\documentstyle[12pt]{article}
\def\laq{\raise 0.4 ex \hbox{$<$}\kern -0.8 em\lower 0.62 ex\hbox{$\sim$}}
\def\gaq{\raise 0.4 ex \hbox{$>$}\kern -0.7 em\lower 0.62 ex\hbox{$\sim$}}

\input{epsf}

\def\APP{{\it Acta Phys. Pol.} }

\def\CQG{{\it Class. Quantum Gravity} }

\def\IJMP{{\it Int. J. Mod. Phys.} }

\def\PL{{\it Phys. Lett.} }
\def\PR{{\it Phys. Rev.} }
\def\PRL{{\it Phys. Rev. Lett.} }

 \def\frac#1#2{{\textstyle{{#1}\over
{#2}}}} 
\def\lsim{\mathrel{\rlap{\lower4pt\hbox{\hskip1pt$\sim$}}
    \raise1pt\hbox{$<$}}} \def\gsim{\mathrel{\rlap{\lower4pt\hbox{\hskip1pt$\sim$}}
    \raise1pt\hbox{$>$}}}
\def\sqr#1#2{{\vcenter{\vbox{\hrule height.#2pt
         \hbox{\vrule width.#2pt height#1pt \kern#1pt
         \vrule width.#2pt}
         \hrule height.#2pt}}}} 

\def\beq{\begin{equation}}
\def\eeq{\end{equation}}
\def\beqa{\begin{eqnarray}}
\def\eeqa{\end{eqnarray}}

\begin{document}

\begin{flushright}
{November 2004} \\
\end{flushright}
\vglue 1cm

\begin{center}
{{\bf  Pioneer's Final Riddle *}}

\vglue 0.5cm
{O.\ Bertolami and J. P\'aramos}

\vglue 0.2cm

{E-mail addresses: {\tt orfeu@cosmos.ist.utl.pt; x\_jorge@netcabo.pt}}

\bigskip
{\it Instituto Superior T\'ecnico,
Departamento de F\'\i sica,\\}
\smallskip
{\it Av.\ Rovisco Pais, 1049-001 Lisboa, Portugal\\}
\end{center}

\setlength{\baselineskip}{0.7cm}

\vglue 0.7cm

\centerline{{\bf Abstract}}

\vglue 0.7cm

\noindent
Launched more than thirty years ago and now drifting in
space with no further contact, the Pioneer 10 and 11 spacecraft
are currently at the center of a small but developing concern: are
they under the influence of an anomalous acceleration that
reflects new, yet unknown physics, or merely experiencing some
unexpected technical problem ? The proposals vary from basic
engineering principles to extra spatial dimensions, but most
probably an answer can only be obtained after a dedicated mission
is underway.

\vfill

\noindent
{\bf *} Talk presented by O.B. at the XIV Encontro Nacional de Astronomia 
e Astrof\'\i sica, Angra do Hero\'\i smo, A\c cores, 22-23 July 2004. 

\vglue 0.7cm

\pagestyle{plain}

\setcounter{equation}{0}
\setlength{\baselineskip}{0.7cm}

\newpage

\begin{figure}
\epsfysize=10cm \epsffile{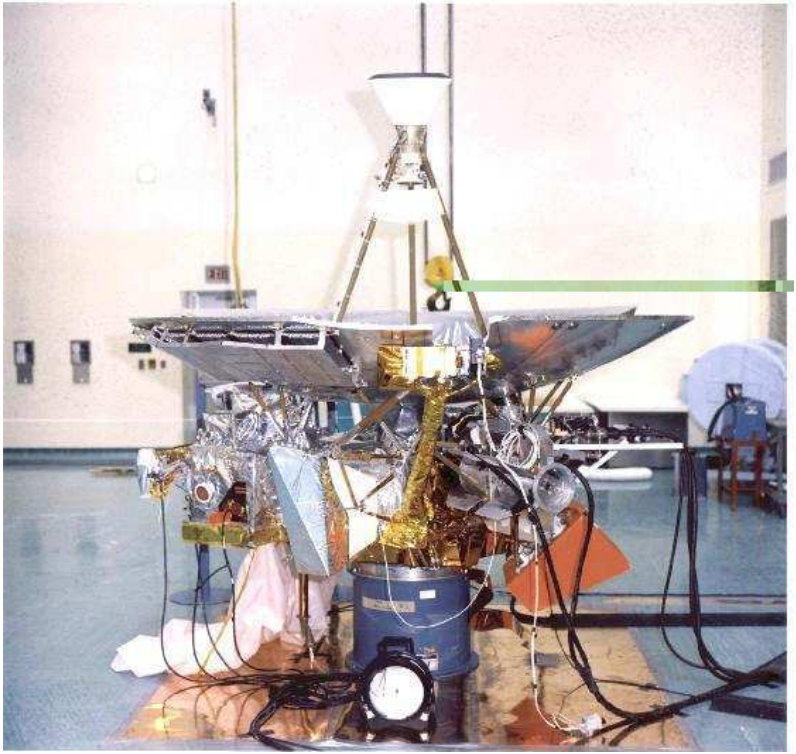} \caption{The Pioneer 10
spacecraft before launch} \label{pioneer}
\end{figure}

\begin{figure}
\epsfysize=10cm \epsffile{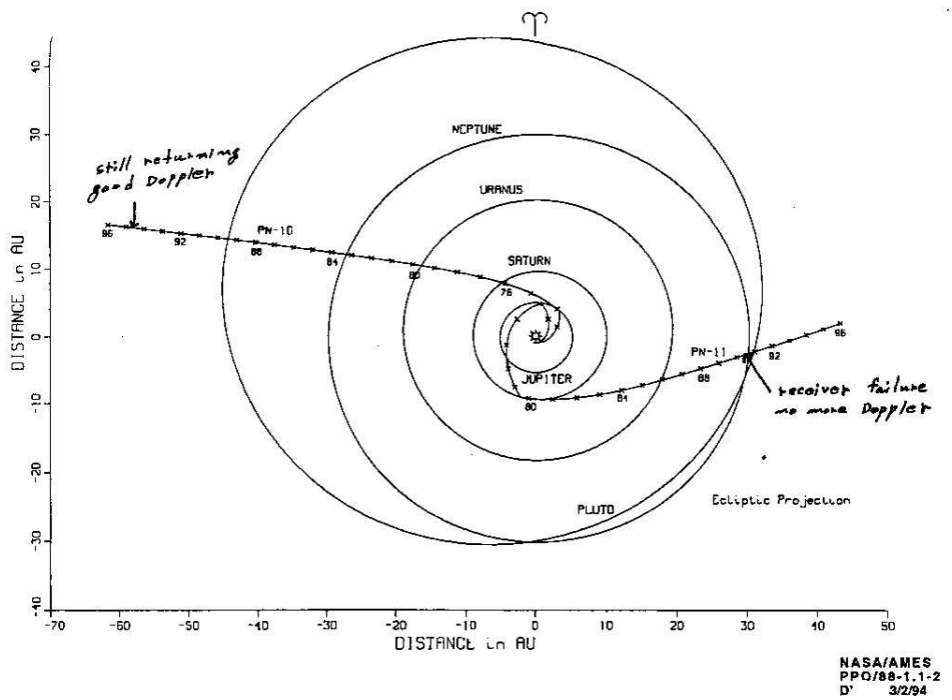} \caption{Trajectories of
Pioneer 10 and 11} \label{trajectories}
\end{figure}

More than thirty years ago, on 2 March 1972, Pioneer 10 was
launched by an Atlas/Centaur rocket from Cape Canaveral, in a
voyage to the yet unchallenged far ends of the Solar System (Figure
\ref{pioneer}). About a year later, on 5 April 1973, Pioneer 11
would follow. In their epic voyage Pioneer 10 would encounter
Jupiter, while Pioneer 11 would also visit Saturn, returning
detailed pictures of the planets and their moons. The two
spacecraft then followed hyperbolic orbits near the plane of the
ecliptic to opposite sides of the Solar System (Figure
\ref{trajectories}). Initial projections gave the spacecraft a
lifespan of about seven years, after which contemporary deep space
scanning stations would no longer be able to determine its
position and accurately receive scientific data. However, further
developments in tracking capabilities extended the probes' life to
an ever increasing limit. Hence, although the Pioneer 10 mission
officially ended on 31 March 1997, contact was kept until
7 February, 2003, when it was at about 70 Astronomical Units from Earth (1
AU is the medium Earth-Sun distance, about 150 million
kilometers). A radio failure on 1 October 1995 and the increasing
exhaustion of the power source rendered the Pioneer 11 unable to
contact us, when it was at approximately 40 AUs from our planet.

After flying by the outer planets they so successfully observed,
both missions were scaled down. In the course of events, tracking
of the spacecraft became a routine training exercise for future
Lunar Prospector controllers. However, it was during this less
splendorous part of their voyage that the Pioneer 10 and 11 cast
their final riddle, which is currently known as the Pioneer
anomaly. Analysis of radiometric data from the Pioneer 10/11 by
NASA's Jet Propulsion Lab \cite{Anderson1} has revealed the
existence of an anomalous acceleration on both spacecraft,
inbound to the Sun and with a (constant) magnitude of $a_A \simeq
(8.5 \pm 1.3) \times 10^{-10}~ m~s^{-2}$. This is not expected
from the usual dynamics, which account solely for the
gravitational pull and the outward solar wind pressure, both
varying with the distance to the Sun $r$ according to the usual
inverse square law $1/r^2$. It was measured from 1980 on, when
Pioneer 10 was at a distance of 20 AUs from the Sun and the solar
radiation pressure acceleration on Pioneer 10 had decreased to
below $5 \times 10^{-8}~m~s^{-2}$. This was possible because the
Pioneer spacecraft were excellent for dynamical studies due to
their spin-stabilization and their great distances, requiring
a minimum number of Earth-attitude reorientation maneuvers for
these deep space missions to go beyond the Solar System.
Contrariwise, the Voyager spacecraft were not suited for precise
celestial mechanics experiments, as numerous attitude-control
maneuvers overwhelm any small external acceleration. This Pioneer
effect seems also to have disturbed the Ulysses and Galileo spacecraft.

As a first attempt to explain this phenomena, the team who
unraveled the anomaly resorted to poor accounting of thermal and
mechanical effects, as well as errors in the tracking algorithms
used. However, these attempts have all failed to account for a
constant acceleration. Indeed, thermal effects such as gas leaks,
heat radiation due to the two nuclear generators placed on booms
or to the instruments on-board all obey exponential laws.
In the same way as air will quickly escape a balloon while it is
full, but decrease its velocity as it is depleted, all heat
radiation decreases exponentially with time, towards zero. Hence,
even if a lucky combination of effects could result in an
acceleration inbound the Sun, in principle it could not render a
constant one (see Ref. \cite{Scheffer} for a counter claim).

The algorithms used to calculate the acceleration of the
spacecraft make use of the Doppler effect observed in the radio
signal received from the spacecraft, much in the same way as one
can perceive if a car is approaching or distancing from the
increase or decrease of its sound pitch. A detailed analysis of
the numerical conundrums of these algorithms also fails to account
for an anomalous acceleration \cite{Anderson2}. Hence, all evidence
points that the Pioneer anomaly is a real effect, not a numerical
contraption due to lack of precision.

There is a stronger argument against a conventional engineering
explanation of the anomalous acceleration the spacecraft were
subjected to. While both Pioneer 10 and 11 were completely equal
in design, the Galileo spacecraft, designed to explore the Jovian
system, and the joint ESA-NASA Ulysses heliosphere observatory are
very different in nature. Also, the two Pioneer spacecraft follow
approximate opposite hyperbolic trajectories away from the Solar
System, while Galileo and Ulysses describe closed orbits.

Given this, it seems unreasonable that three different designs and
trajectories would all suffer different conventional effects such
as heat radiation and gas leaks, but originating the same net result.
Remember that the anomalous acceleration was found not only to be
constant and inbound to the Sun, but also textit{similar} for all
four spacecraft.

Thus, it is plausible to assume that there is no conventional
explanation for the Pioneer anomaly and one must resort to new
physics to account for it. From a theoretical point of view, the
phenomena is puzzling: one has to device a theory which breaks 
the usual law of gravity in a very subtle way, so to affect only the
motion of small bodies such as the referred spacecraft, for there
is no analogous effect for planets. Hence, one of the very
foundations of General Relativity is at stake: the Weak Principle
of Equivalence, which states that all bodies fall in a
gravitational field at the same rate, independently of their mass
or constitution.

Physicists are, of course, happy to come across evidence of new physics.
Thus, a wide range of more or less exotic hypotheses has been put forward to
meet the challenge. Some propose a new interaction \cite{Anderson2}, or
recover modifications of gravity suggested as alternatives to
dark matter \cite{Milgrom} (see also \cite{Bertolami1}),
or seek for gravity models
in which the Newtonian dynamics do not follow in the weak-field limit 
\cite{Cadoni}, or consider 
a scalar field with a suitable potential \cite{Bertolami2}.
Another suggestion considers the dependence on the momentum of the 
gravitational coupling, the main issue in 
Refs. \cite{Bertolami1,Goldman} and, as pointed in Ref. \cite{Bertolami2}, a 
potential solution for the anomaly, in the context of the 
linear approximation of gravity and, 
interestingly, two potentials: the usual Newtonian one, and a second that 
generalizes the PPN parameter $\gamma$ to 
a function of distance \cite{Jaekel-Reynaud}. Given that the value of the
anomalous acceleration can be related to those of the speed of
light, $c = 3 \times 10^8~m~s^{-1}$, and the Hubble constant 
$H_0 \simeq 70~km~s^{-1}~Mpc^{-1}
\simeq 2.3 \times 10^{-18}~s^{-1}$ (which allows measuring the velocity
of expansion of the Universe and its age) through $a_A = c H_0 $, some authors
argue that the phenomena merely
reflects the ongoing Universe's expansion \cite{Ostvang}. Others claim
that there exists also a connection with the cosmological constant 
\cite{Nottale}, dubbed ``my greatest error'' by Einstein himself, but
confirmed to be non-vanishing by different cosmological observations
\cite{Bahcall}. However, it seems clear that
this should produce an \textit{outgoing} acceleration, in the same
way as dots on an inflating balloon's surface seem to pull away
from each other. Also, this effect should be observed on all
bodies, regardless of their scale, not only on the spacecraft.
Finally, the computed effect due to the cosmological constant is 
much smaller than the anomalous acceleration felt by the
spacecraft.

Another possibility involves the idea of an extra spatial
dimension which is, of course, also appealing for other reasons.
Extra dimensions are a crucial element in unifying models of the
fundamental interactions of Nature. Imagine, for instance, an ant
living on top of a sheet of paper, oblivious of the
three-dimensional world around it; if one turns on a light at the
center of the sheet, the ant would expect its intensity to vary
with the distance to the source, $r$, according to $1/r$. Thus,
the same flux of radiation would transverse circles of different
radius, according to the energy conservation law. However, a
three-dimensional observer such as ourselves would expect the
dependency to be of the form $1/r^2$, since we see the flux
emanating not in circles, but as spheres.

By the same token, we expect gravity to vary as $1/r^2$ because we
live in a world with three spatial dimensions. However, if there
is a fourth spatial dimension and we live on a three-dimensional
sheet - a brane - gravity should behave as $1/r^3$ at large
distances from the source, where this extra ``length'' becomes
noticeable. Conversely, while we still expect gravity to obey the
usual $1/r^2$ Newton's law at small distances, there should be
some small corrections to it.

Such a general picture, which is unrelated to the problem at hand,
is usually referred to as the braneworld scenario \cite{Randall}.
According to it, we live on a three-dimensional brane in a 
five-dimensional space,
with a possible second brane at a distance. The tension between
these branes (which keeps them at a fixed distance, like a rod
between two brick walls) is perceived by us as the cosmological
constant. In the five-dimensional
space away from the brane only gravity is allowed to propagate.

In a violin, vibrations are trapped between two fixed extrema, and
only a certain number of notes are allowed, related to the length
of each chord and the velocity of sound. By analogy, if we have
two branes and gravity propagating between them, it will be
constrained to exhibit certain characteristic modes, limited by
the distance between those and by the speed of light. Also, small
oscillations of one membrane relative to the other could lead to a
change in gravity observed on each brane \cite{Gregory}. Such
effects may be tested some day in the future, but they cannot
account directly for the anomalous constant acceleration the
Pioneer spacecraft are subjected to \cite{Bertolami2}.
Nevertheless, a more elaborate argument involving the presence of
a scalar field with an appropriate potential still allows for an
explanation of the Pioneer anomaly \cite{Bertolami2}.

\begin{figure}
\epsfysize=10cm \epsffile{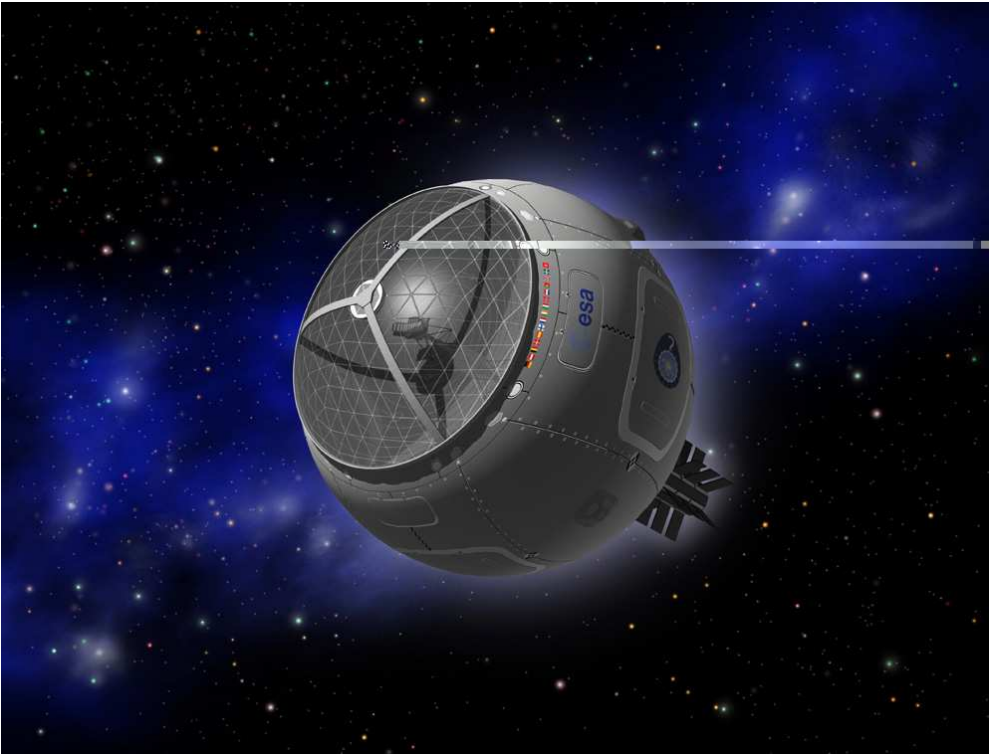} \caption{Artist's
conception of the Sputnik V mission} \label{sputnik}
\end{figure}

However, knowledge does not advance solely from \textit{ad-hoc}
hypotheses with the purpose of obtaining the correct figures, or
even extracting yet unforeseen results from well established
theories. A fundamental aspect of any scientific observation of a
phenomena is its replication. Without a bulletproof confirmation
by a dedicated experiment, no conclusions can be drawn. With this
in mind, a mission designed solely with the purpose of accurately
measuring its own motion has been proposed by two groups
independently \cite{Bertolami3,Anderson3}; in a sense, a
sophisticated pebble traversing space. In one of the proposals,
its shape should be as symmetric as possible, so to exclude any
anisotropic effects; engines and energy sources
should be as far as possible from
the main body, to reduce vibrations and heat dissipation effects that 
might affect readings;
and refined differential accelerometers should be on board, providing a direct,
mechanical measurement of the probes' motion, together with the
usual Doppler tracking. Such a mission has been dubbed Sputnik V,
recalling the first-ever satellite and the current objective of
searching for a ``fifth'' new interaction of Nature
\cite{Bertolami3,Bertolami4,Bertolami5} (Figure \ref{sputnik}). Alternative
mission concepts have also been discussed \cite{Anderson3,Turyshev}.

Although perhaps initially viewed with some suspicion and
skewness, the study of the Pioneer anomaly has been steadily
gaining momentum, as can be easily seen from the number of related
peer reviewed papers published in the specialized literature. The
realization of the importance of this unpredicted phenomenum as a
natural testing ground for space science and 
gravitation has reached its climax with
the first dedicated conference, held last May at Bremen's
University Center for Applied Space Technology and Microgravity
(ZARM). There, researchers around the world converged to
discuss the subject, with backgrounds ranging from pure
engineering to applied and fundamental physics.

The discussion of the abovementioned mission sketches was one of
the topics of discussion, leading to the concatenation of efforts so to
supply the European Space Agency with proposals for the Cosmic
Vision 2015-2025 call \cite{Cosmicvision1}. This effort has yielded five 
different mission concepts \cite{Cosmicvision2}, 
among over 150 proposed in all subjects, 
and ongoing efforts to merge them into an unified one.

If ever approved to fly, this mission will develop new and exciting 
deep space navigational concepts and hopefully uncover new 
physics. A confirmation of the Pioneer
anomaly might unfold interesting theoretical thinking, which may
have an impact on the way we regard the history of
the Universe as a whole.

\vfill
\newpage

\end{document}